\documentstyle[11pt,a4,]{article}

\begin{document}

\title{Diagonalization of $pp$-waves}
\author{B.V.Ivanov\thanks{%
E-mail address: boyko@inrne.acad.bg} \\
Institute for Nuclear Research and Nuclear Energy\\
Tzarigradsko Shausse 72, Sofia 1784, Bulgaria}
\date{30 April 1997}
\maketitle

\begin{abstract}
A coordinate transformation is found which diagonalizes the axisymmetric $pp$%
-waves. Its effect upon concrete solutions, including impulsive and shock
waves, is discussed.

PACS No. 04.20J
\end{abstract}
\newpage\

\section{Introduction}

A well-known class of gravitational waves are the plane-fronted waves with
parallel rays ( $pp$-waves) which admit a covariantly constant null vector 
\cite{one},\cite{two},\cite{three},\cite{four}. The metric can be written
using cylindrical coordinates: 
\begin{equation}
ds^2=2dudW+2Hdu^2-dP^2-P^2d\Phi ^2  \label{one}
\end{equation}
where $H=H(u,P,\Phi )$ and is of Petrov type $N$ or conformally flat. Mainly
axisymmetric $pp$-waves have been discussed in the literature. Such is the
Schwarzschild solution boosted to the speed of light, which is interpreted
as a massless null particle \cite{five},\cite{six} or an ultrarelativistic
black hole \cite{seven},\cite{eight},\cite{nine}. A ring of massless
particles is produced by boosting the Kerr metric \cite{ten}. Axisymmetric $%
pp$-waves describe also the gravitational field of light beams \cite{eleven},%
\cite{twelve}. Plane-fronted electromagnetic waves generate $pp$%
-gravitational waves as exact solutions of the Einstein-Maxwell equations 
\cite{thirteen},\cite{fourteen},\cite{fifteen}.

While superposition of $pp$-waves running in the same direction is trivial,
their collisions have been studied mainly for the subclass of plane waves 
\cite{sixteen},\cite{seventeen} in which the physical invariants are
constant over the wave surfaces. The reason is that (1) is unsuitable to
describe two approaching waves, but in the case of plane waves a
transformation due to Rosen \cite{eighteen} converts (1) into the Szekeres
line element \cite{four},\cite{seventeen},\cite{nineteen} which can
encompass two approaching waves and the region of their interaction. For a
wave of constant polarization it reads 
\begin{equation}
ds^2=2dudv-e^{-U}\left( e^Vdr^2+e^{-V}d\varphi ^2\right)  \label{two}
\end{equation}
where $U$ and $V$ are functions of the null coordinate $u$ . Having in mind
that $\sqrt{2}u=t-z$, $\sqrt{2}v=t+z$ it is clear that (2) is a diagonal
metric. Unfortunately, such waves have infinite extent and energy. The waves
given by (1) with a suitably chosen $H$ are finite in extent and energy but
are not asymptotically flat and contain metric discontinuities for impulsive
and shock waves.

In the present paper we generalize the Rosen transformation to axisymmetric $%
pp$-waves and find for them a diagonal and asymptotically flat form like
(2). This may be considered both as an alternative description and as a
preparatory step before the investigation of their head-on collisions.

\section{Generalized Rosen transformation}

Let us investigate upon what conditions on $H(u,P,\Phi )$ the metric (1) can
be diagonalized. Following Rosen, we do not change $u$ at all. The
requirement $g_{uv}=1$ , like in (2), is ensured by $W=v+W_1\left(
u,r,\varphi \right) $ . Next, the vanishing of $g_{vv}$ takes place when $%
P_v=\Phi _v=0$ . The non-diagonal term $g_{r\varphi }$ disappears if 
\begin{equation}
P_rP_\varphi +P^2\Phi _r\Phi _\varphi =0  \label{three}
\end{equation}
Obviously $P_r\neq 0$ , $\Phi _\varphi \neq 0$ and the simplest solution of
(3) is $P_\varphi =\Phi _r=0$ . Then the remaining non-diagonal terms
disappear when 
\begin{equation}
\begin{array}{llll}
W_{1r}=P_uP_r &  &  & W_{1\varphi }=P^2\Phi _u\Phi _\varphi
\end{array}
\label{four}
\end{equation}
with $P\left( u,r\right) $ and $\Phi \left( u,\varphi \right) $ .
Integrating the first equation in (4) and plugging it into the second we
find a l.h.s. independent of $r$ , unlike the r.h.s. The simplest solution
is $\Phi _u=0$ and $\Phi =\varphi $ which does not change the range of the
angular coordinate. Hence $W_1=W_1\left( u,r\right) $ . The nullification of 
$g_{uu}$ is guaranteed by the relation 
\begin{equation}
2H=-2W_{1u}+P_u^2  \label{five}
\end{equation}
The r.h.s. depends on $u$ and $r$ , therefore after the coordinate
transformation $H=H\left( u,r\right) $, which means that in the beginning
it was $H\left( u,P\right) $ . Thus the sufficient condition for the
diagonalization of (1) is the axial symmetry of $H$ . We can't say that it
is also necessary because the simplest solutions of (3,4) have been used.

Integrating eq(4), inserting the result into (5) and taking the $r$%
-derivative we obtain the main equation which governs the generalized Rosen
transformation: 
\begin{equation}
P\left( u,r\right) _{uu}=-H\left( u,P\right) _P  \label{six}
\end{equation}

The line element (1) becomes 
\begin{equation}
ds^2=2dudv-P_r^2dr^2-P^2d\varphi ^2  \label{seven}
\end{equation}
which is diagonal and the coordinate transformation reads 
\begin{equation}
\begin{array}{lllllll}
u=u &  & \Phi =\varphi &  & P=P\left( u,r\right) &  & W=v+W_1\left(
u,r\right)
\end{array}
\label{eight}
\end{equation}
where $P$ is determined by (6) and $W_1$ by (4). $W$ does not appear in
(6,7).

Usually (1) is written in cartesian coordinates: 
\begin{equation}
ds^2=2dudW+2H\left( u,X,Y\right) du^2-dX^2-dY^2  \label{nine}
\end{equation}
For comparison, we apply the same diagonalization process to (9) with the
following results. $H$ must be separable 
\begin{equation}
H=H_1\left( u,X\right) +H_2\left( u,Y\right)  \label{ten}
\end{equation}
there are two main equations 
\begin{equation}
\begin{array}{llll}
X\left( u,x\right) _{uu}=-H_1\left( u,X\right) _X &  &  & Y\left( u,y\right)
_{uu}=-H_2\left( u,Y\right) _Y
\end{array}
\label{eleven}
\end{equation}
the line element becomes 
\begin{equation}
ds^2=2dudv-X_x^2dx^2-Y_y^2dy^2  \label{tweleve}
\end{equation}
and the coordinate transformation reads 
\[
\begin{array}{lllllll}
u=u &  &  & X=X\left( u,x\right) &  &  & Y=Y\left( u,y\right)
\end{array}
\]
\begin{equation}
W=v+\int^xX_uX_{x^{\prime }}dx^{\prime }+\int^yY_uY_{y^{\prime }}dy^{\prime }
\label{thirteen}
\end{equation}

In vacuum the only surviving Einstein equation gives 
\begin{equation}
H_{1XX}+H_{2YY}=0  \label{fourteen}
\end{equation}
After the usual removal of linear terms in $H_i$ (11) becomes linear: 
\begin{equation}
\begin{array}{llll}
X_{uu}=-f\left( u\right) X &  &  & Y=f\left( u\right) Y
\end{array}
\label{fifteen}
\end{equation}
where $f\left( u\right) $ is the second derivative of $H_1$ . This permits
the choice $X=xF\left( u\right) $, $Y=yG\left( u\right) $ which represents
exactly the Rosen transformation for plane waves of constant polarization 
\cite{four},\cite{eighteen},\cite{twenty}. Thus in vacuum the
diagonalization of $pp$-waves in cartesian coordinates requires separability
(10) and leads directly to plane waves.

\section{$pp$-waves in Brinkmann and diagonal form}

We have shown that an axisymmetric $pp$-wave can be described by (7). At
first sight (7) is a special case of the Szekeres line element (2) where now 
$U$ and $V$ depend on $u$ and $r$. In fact, there is no loss of generality
and any metric (2) may be written as (7) provided the Einstein equations
hold. Let us compare these equations for (1,7) and (2).

A $pp$-wave allows energy-momentum tensors of few types: vacuum, null
electromagnetic field or pure radiation (null dust), which may be combined 
\cite{fourteen}, \cite{fifteen} . All of them have only one non-trivial
component: 
\begin{equation}
T_{uu}=2\rho \left( u,P\right) =2\rho _R+2\rho _E  \label{sixteen}
\end{equation}
where $\rho _E$ is the energy-density of pure radiation with no matter
equations and $\rho _E$ is the electromagnetic energy-density 
\begin{equation}
\begin{array}{llll}
2\rho _E=\nabla \psi \nabla \psi &  &  & \psi =A_u
\end{array}
\label{seventeen}
\end{equation}
We have used relativistic units with $8\pi G/c^4=1$ . $A_u$ is the only
component of the vector potential and satisfies the Maxwell equation $%
\triangle \psi =0$ . We suppose that no charges and currents are present.
The gradient and Laplacian are with respect to $P$, $\Phi $ (or $X$, $Y$ ).
We have chosen this formalism instead of the Newman-Penrose one since $%
T_{uu},$ $A_u$ and the Maxwell equation do not change under the generalized
Rosen transformation (8).

In Brinkmann coordinates the only non-trivial Ricci tensor component is $%
R_{uu}$ which is a Laplacian. The Einstein equation is 
\begin{equation}
H_{PP}+\frac 1PH_P=2\rho  \label{eighteen}
\end{equation}
and its solution is well-known from classical potential theory: 
\begin{equation}
H_P=\frac 2P\int_0^P\rho \left( u,P^{\prime }\right) P^{\prime }dP^{\prime
}+\frac 2P\rho _e\left( u\right)  \label{nineteen}
\end{equation}

\begin{equation}
H=2\int_0^P\frac{dP^{\prime }}{P^{\prime }}\int_0^{P^{\prime }}\rho \left(
u,P^{\prime \prime }\right) P^{\prime \prime }dP^{\prime \prime }+2\rho
_e\left( u\right) \ln \frac Pa  \label{twenty}
\end{equation}
where an ignorable term has been omitted in (20). The second term in (19,20)
is the exterior solution with arbitrary $\rho _e$ and some constant length $%
a $ . The Maxwell equation is 
\begin{equation}
\left( P\psi _P\right) _P+\frac 1P\psi _{\varphi \varphi }=0
\label{twentyone}
\end{equation}
We have retained some $\varphi $-dependence in $\psi $ but it must disappear
in $\rho _E$ . The only non-zero Weyl scalar is given by 
\begin{equation}
\Psi _4=\rho -\frac{H_P}P  \label{twentytwo}
\end{equation}

Now, let us concentrate on the metric (2) with $U\left( u,r\right) $ , $%
V\left( u,r\right) $ . The Einstein equations yield: 
\begin{equation}
2U_{uu}=U_u^2+V_u^2+4\rho  \label{twentythree}
\end{equation}
\begin{equation}
\left( U+V\right) _{ru}=\left( U+V\right) _rV_u  \label{twentyfour}
\end{equation}
\begin{equation}
\left( U+V\right) _{rr}=\left( U+V\right) _rV_r  \label{twentyfive}
\end{equation}
Eqs(24,25) are easily integrated 
\begin{equation}
\left( U+V\right) _r=-2e^V  \label{twentysix}
\end{equation}
We have chosen the integration constant in order to restore Minkowski
spacetime for $V=U$ , $r=e^{-V}.$ This allows smooth transition to it in
front of the wave and is in accord with our demand for asymptotically flat
solutions. The Maxwell equation is 
\begin{equation}
\left( e^{-U}\psi _r\right) _r+e^V\psi _{\varphi \varphi }=0
\label{twentyseven}
\end{equation}
where like in (21) we allow for some $\varphi $-dependence. Using the
natural NP null tetrad \cite{four} we find three non-trivial Weyl scalars: 
\begin{equation}
\begin{array}{llll}
\Psi _2=-\frac 1{12}e^{U-V}\left[ \left( U+V\right) _{rr}-\left( U+V\right)
_rV_r\right] &  &  & \Psi _3=-\frac{\sqrt{2}}4e^{\frac{U-V}2}R_{ur}
\end{array}
\label{twentyeight}
\end{equation}
\begin{equation}
\Psi _4=\frac 12\left( V_uU_u-V_{uu}\right)  \label{twentynine}
\end{equation}
A look at the Einstein equations (23-25) shows that $\Psi _2=\Psi _3=0$ and
the field is of type $N$ not of type II.

One can try to solve the relevant eqs(23,26), when $\rho $ is given , in two
ways. First, we may take an arbitrary $U$ , solve for $V$ from (23) and
insert the result into (26). This leads to a condition on $U$ : 
\begin{equation}
\begin{array}{llll}
\left( U+A\right) _r=a_1\left( r\right) e^A+a_2\left( r\right) &  &  & 
A=\int \sqrt{2U_{uu}-U_u^2-4\rho }du
\end{array}
\label{thirty}
\end{equation}
with arbitrary $a_1\neq 0$ and $a_2$ . Eq(30) is too complicated to be
examined. Second, we take an arbitrary $V$ and notice that (23) is a linear
second-order equation for $e^{-U/2}$ . Then we integrate (26): 
\begin{equation}
U=-\int \left( 2e^V+V_r\right) dr+f_1\left( u\right)  \label{thirtyone}
\end{equation}
where $f_1\left( u\right) $ is some yet undetermined function. Substituting
(31) into (23) we get an equation for $f_1$ with additional conditions on $V$
to yield a $r$-independent $f_1$ which again are very complicated.

Let us now compare the two theories. The link is given by (7): 
\begin{equation}
\begin{array}{llll}
P=e^{-\frac{U+V}2} &  &  & P_r=e^{\frac{V-U}2}
\end{array}
\label{thirtytwo}
\end{equation}

Eq(32) gives at first sight an additional constraint between $U$ and $V$ but
this turns out to be exactly (26) which is necessarily satisfied. Going
backwards, (26) shows that (32) holds for some $P$ . It can be shown further
that (6,18) are equivalent via (32) to (23). The same is true about (21) and
(27) which is not so surprising for a Laplacian. At last, under (32) the
Weyl scalar (22) coincides with (29). Consequently, axisymmetric $pp$-waves
(1) are in a one-to-one correspondence with the solutions $U\left(
u,r\right) $, $V\left( u,r\right) $ for metric (2). Thus we can replace
metric (2) with two functions by metric (7) with one function $P$ or by
metric (1) with one function $H$ . Each of these forms has its own merits.
The Brinkmann metric (1) has simple Einstein equations (compare (18) with
(30,31)) but is not asymptotically flat for exterior solutions, sometimes
has a discontinuous $H$ and is unfit for studying collisions of $pp$-waves
because of only one null coordinate. The metric given by (7) or (2,32) is
worth as a starting point for the interaction problem and is asymptotically
flat for realistic $\rho $ as will be shown in the following. However, $%
0\leq P\leq \infty $ as a radial coordinate in (1) which makes $g_{\varphi
\varphi }$ in (7) singular at some points. This coordinate singularity is
innocuous when it is due to the cylindrical character of the coordinate
system. If not, the experience with plane waves teaches that it becomes a
fold singularity and is intimately related to the curvature singularity in
the interaction region \cite{twentyone}.

\section{Solutions: general features}

Eq(6) with $H$ satisfying (20) is a second-order nonlinear differential
equation with respect to $P$ . In the process of solving it arbitrary
functions of $r$ arise which reflect the residual freedom in the coordinate
transformation and may be selected to further simplify the solution and
satisfy boundary conditions.

The trivial Minkowski solution is given by $H=0$ , $P=r$ , $P_r=1$ . Having
in mind the setting of the collision problem, $u=0$ must be the boundary
between the running wave ( $u>0$ ) and Minkowski spacetime ( $u<0$ ) where
the wave has not yet arrived. This gives the universal boundary condition 
\begin{equation}
P\left( 0,r\right) =r  \label{thirtythree}
\end{equation}
We also demand asymptotic flatness i.e. $P\left( u,\infty \right) =r$ for
fixed $u$ .

It is clear from (20) that the exterior solution is always separable, the
interior is separable when $\rho =\rho _1\left( P\right) \rho _2\left(
u\right) $ . Almost all $pp$-waves discussed in the literature are of this
kind with $H\left( u,P\right) =H_1\left( P\right) \rho _2\left( u\right) $
and we shall consider only them in the following. A natural question arises:
when $H$ is static in (1) is it possible that $P$ is also static in (7)?
This is not allowed by (4,5). A static $P$ has $P_u=0$ , $W_1=W_1\left(
u\right) $ and $H=H\left( u\right) $ contrary to our assumption that $%
H=H\left( P\left( r\right) \right) $ . Hence $P$ depends on $u$ even when $%
\rho _2\left( u\right) =1$ .

For the exterior solution given in (19) eqs(6,22) become 
\begin{equation}
\begin{array}{llll}
P_{uu}=-\frac 2P\rho _e\left( u\right) &  &  & \Psi _4=-\frac 2{P^2}\rho
_e\left( u\right)
\end{array}
\label{thirtyfour}
\end{equation}
Unlike (15), eq(34) is non-linear and we can't get rid of the $r$%
-dependence. For many simple choices of $\rho _e$ (34) falls in the class of
Emden-Fowler equations \cite{twentytwo}. They are quite difficult to solve
and many of them remain non-integrable. For example, when $\rho _e\left(
u\right) =u^n$ the integrable cases are just $n=0;-1;-2$ . For
asymptotically flat solutions (34) shows that $\Psi _4\rightarrow 0$ when $%
r\rightarrow \infty $ .

The general separable interior solution emerges from 
\begin{equation}
P_{uu}=-H_1\left( P\right) _P\rho _2\left( u\right)  \label{thirtyfive}
\end{equation}
and again is reducible in many cases to the Emden-Fowler equations and their
generalizations. The case $\rho _1\left( P\right) =1$ is special. Then (35)
is linear in $P$ and we can use an analog of the ansatz applied after (15),
namely $P\left( u,r\right) =rp\left( u\right) $ : 
\begin{equation}
\begin{array}{llll}
H=\frac 12\left( X^2+Y^2\right) \rho _2\left( u\right) &  &  & \Psi _4=0
\end{array}
\label{thirtysix}
\end{equation}
\begin{equation}
p_{uu}=-\rho _2\left( u\right) p  \label{thirtyseven}
\end{equation}
\begin{equation}
ds^2=2dudv-p\left( u\right) ^2\left( dX^2+dY^2\right)  \label{thirtyeight}
\end{equation}
This is the case of pure radiation with density $\rho _R=\rho _2$ or an
electromagnetic wave with potential, Maxwell scalar and energy-density given
by 
\begin{equation}
\begin{array}{llll}
\psi =a_3\left( u\right) X+a_4\left( u\right) Y &  &  & \Phi _2\left(
u\right) =-\frac 1{\sqrt{2}}\left[ a_3\left( u\right) -ia_4\left( u\right)
\right]
\end{array}
\label{thirtynine}
\end{equation}
\begin{equation}
\rho _E\left( u\right) =\rho _2\left( u\right) =\frac 12\left(
a_3^2+a_4^2\right)  \label{forty}
\end{equation}
where $a_3$ , $a_4$ are arbitrary functions. Obviously $\psi $ depends on $%
\varphi $ while $\rho _E$ does not. The potential satisfies the Maxwell
equation (21). In fact eqs(36-40) represent a plane electromagnetic wave 
\cite{four} and an axisymmetric electromagnetic $pp$-wave at the same time.
There is no pure gravitational wave in addition because $\Psi _4=0$ . The
Ricci scalar $\Phi _{22}=\Phi _2\bar \Phi _2$ is constant over the wave
surface. On the contrary, pure plane gravitational waves can not be
axisymmetric because their $H\sim X^2-Y^2$ which is $\varphi $-dependent.

The case discussed above provides a link between plane and axisymmetric $pp$%
-waves. Even for it, eq(37) is the normal form of the general linear
second-order equation and its general solution is given analytically only if
a non-trivial concrete solution is known. Therefore we are going to discuss
two cases of simple $u$-dependence when the solution of (6) may be found.
These are the impulsive and shock waves.

\section{Impulsive waves}

These are waves with $\rho _2=\delta \left( u\right) $ . Eq(37) may be
integrated with the help of (33): 
\begin{equation}
\begin{array}{llll}
P=r\left( 1-\frac{H_{1r}}ru\right) &  &  & P_r=1-H_{1rr}u
\end{array}
\label{fortyone}
\end{equation}
Eq(22) transforms into 
\begin{equation}
\Psi _4=\left( \rho _1\left( r\right) -\frac{H_{1r}}r\right) \delta \left(
u\right)  \label{fortytwo}
\end{equation}
which clearly demonstrates the impulsive character of the wave. It is seen
from (19) that $H_{1r}>0$ and (41) shows that $P$ always possesses a
coordinate singularity for some $u>0$, different from the cylindrical
singularity at $r=0$ . This is a consequence of the positive energy
condition and the idealized impulsive character of the wave.

For a boosted Schwarzschild solution \cite{five},\cite{six},\cite{nine} $%
H=2\mu \delta \left( u\right) \ln P^2$ where $\mu $ is the momentum of the
null point-like particle and (41,42) give 
\begin{equation}
ds^2=2dudv-\left[ 1+\frac{4\mu }{r^2}u\theta \left( u\right) \right]
^2dr^2-\left[ 1-\frac{4\mu }{r^2}u\theta \left( u\right) \right]
^2r^2d\varphi ^2  \label{fortythree}
\end{equation}
\begin{equation}
\Psi _4=\left( \delta \left( r\right) -\frac 4{r^2}\right) \mu \delta \left(
u\right)  \label{fortyfour}
\end{equation}
Eq(43) is exactly the line element found in \cite{seven},\cite{eight}. There
is a curvature singularity at the point of the source $r=0$ . $H$ is also
the function for an exterior impulsive solution given in (20). If $t$ is
fixed, for any $z$ and $r\rightarrow \infty $ the solution is asymptotically
flat. There is a coordinate singularity at $\sqrt{2}r^2=4\mu \left(
t-z\right) $ . For a fixed $z$ , as time goes by, the singular circle
centred at $z$ expands towards infinity.

For a boosted Kerr solution \cite{ten}: 
\begin{equation}
H=2\mu \delta \left( u\right) \ln \left| P^2-b^2\right|  \label{fortyfive}
\end{equation}
According to (41): 
\begin{equation}
\begin{array}{llll}
P=r\left( 1-\frac{4\mu }{r^2-b^2}u\right) &  &  & P_r=1+4\mu u\frac{r^2+b^2}{%
\left( r^2-b^2\right) ^2}
\end{array}
\label{fortysix}
\end{equation}
where $b$ is the radius of the ring of massless particles. The curvature
singularity moves to $r=b$ and the region $r\leq b$ is free of coordinate
singularities. The metric is asymptotically flat.

As a final example we present the diagonalization of an impulsive beam of
light with transverse radius $a$ \cite{eleven},\cite{twelve}. This is a
global solution the interior being given by (36) and the exterior by (20).
The junction conditions require that 
\begin{equation}
H=\frac{4mP^2}{a^2}\theta \left( a-P\right) \delta \left( u\right) +4m\left(
1+2\ln \frac Pa\right) \theta \left( P-a\right) \delta \left( u\right)
\label{fortyseven}
\end{equation}
where $m$ is the constant energy density. With the help of (18) eq(41) may
be rewritten as 
\begin{equation}
\begin{array}{llll}
P=r\left( 1-\frac{H_{1r}}ru\right) &  &  & P_r=1+\left( \frac{H_{1r}}r-2\rho
_1\right) u
\end{array}
\label{fortyeight}
\end{equation}
Inserting (47) into (48) we obtain for the interior and exterior solutions: 
\begin{equation}
\begin{array}{llll}
P_i=r\left( 1-\frac{8m}{a^2}u\right) &  &  & P_{ir}=1-\frac{8m}{a^2}u
\end{array}
\label{fortynine}
\end{equation}
\begin{equation}
\begin{array}{llll}
P_e=r\left( 1-\frac{8m}{r^2}u\right) &  &  & P_{er}=1+\frac{8m}{r^2}u
\end{array}
\label{fifty}
\end{equation}
It is seen that $P$ is continuous at $r=a$ but $P_r$ makes a finite jump.
According to (48) the reason is the jump in $\rho _1$ from zero to a finite
constant, since the junction conditions require that $H_1$ and $H_{1r}$
should be continuous. Consequently, solutions which are perfectly well
joined in Brinkmann coordinates acquire discontinuous metric upon
diagonalization due to unrealistic densities with $\theta \left( r\right) $
terms. The problem disappears when the density smoothly falls to zero. Take
for example $\rho _1\left( P\right) =e^{-P^2}$ . Then 
\begin{equation}
H=\left[ \ln P-\frac{1}{2}{\rm Ei}\left( -P^2\right) \right] \delta \left(
u\right)  \label{fiftyone}
\end{equation}
\begin{equation}
\begin{array}{llll}
P=r\left( 1-\frac{1-e^{-r^2}}{r^2}u\right) &  &  & P_r=1+\frac{%
1-e^{-r^2}-2r^2e^{-r^2}}{r^2}u
\end{array}
\label{fiftytwo}
\end{equation}
When $P\rightarrow 0,\infty $ $H$ in (51) approaches the first or the second
term in (47). Correspondingly, when $r\rightarrow 0$ (52) approaches (49)
and when $r\rightarrow \infty $ it approaches (50) with $8m=a=1$ . The
metric (52) is asymptotically flat but the coordinate singularities still
exist.

\section{Shock waves}

These waves have $H=H_1\left( P\right) \theta \left( u\right) $ and (6) has
a first integral: 
\begin{equation}
P_u^2=c\left( r\right) -2H_1\left( P\right)  \label{fiftythree}
\end{equation}
It is clear from (20) that $H_1$ is a positive and increasing function. This
is the reason to keep the arbitrary function $c\left( r\right) $ in (53) so
that the r.h.s. is positive. Eq(53) is easily integrated. Imposing (33) we
obtain 
\begin{equation}
\pm u=\int_r^P\frac{dP^{\prime }}{\sqrt{c\left( r\right) -2H_1\left(
P^{\prime }\right) }}=K\left( P,r\right) -K\left( r\right)  \label{fiftyfour}
\end{equation}
which gives $P\left( u,r\right) $ indirectly. For future convenience we have
introduced also the indefinite integral $K$ .

In order to understand the meaning of $c\left( r\right) $ let us discuss the
interior solution (36-38) with $\rho _2\left( u\right) =\theta \left(
u\right) =1$ in the region occupied by the wave. The integral in (54) can be
evaluated: 
\begin{equation}
\arcsin \frac P{\sqrt{c}}=-u+\arcsin \frac r{\sqrt{c}}  \label{fiftyfive}
\end{equation}
and $P$ is found by inverting (55). Let us choose 
\begin{equation}
c\left( r\right) =2H_1\left( r\right)  \label{fiftysix}
\end{equation}
Then we obtain 
\begin{equation}
P=r\cos u  \label{fiftyseven}
\end{equation}
\begin{equation}
ds^2=2dudv-\cos ^2u\ \left( dr^2+r^2d\varphi ^2\right)  \label{fiftyeight}
\end{equation}
This, however, is the line element of an electromagnetic shock wave with
Ricci scalar $\Phi _{22}=\theta \left( u\right) $ \cite{four} and this is
really the case here because we can choose in (39) $a_3=\sqrt{2}\theta
\left( u\right) $, $a_4=0$, $\Phi _2=-\theta \left( u\right) $ . We
conclude that plane waves recommend the receipt (56). Eq(57) shows that it
is equivalent to the method used in (35-37) for the linear case.

Let us apply this receipt to the exterior solution $H_1\left( P\right) =b\ln
\frac Pa$ , $b>0$ . The integral in (54) yields the error function 
\begin{equation}
{\rm erf}\sqrt{\ln \frac rP}=\sqrt{\frac{2b}\pi }\frac ur
\label{fiftynine}
\end{equation}
This formula may be inverted 
\begin{equation}
P=r\exp \left\{ -\left[ {\rm erf}^{-1}\left( \sqrt{\frac{2b}\pi }\frac
ur\right) \right] ^2\right\}  \label{sixty}
\end{equation}
The metric satisfies the necessary boundary condition (33) and is
asymptotically flat for fixed $u$ . The problem is that ${\rm erf}z\leq
1 $ and (59) imposes the constraint 
\begin{equation}
u\leq \sqrt{\frac \pi {2b}}r  \label{sixtyone}
\end{equation}
The solution (60) does not cover the whole region $u\geq 0$ , $0\leq r\leq
\infty $ .

This is a generic feature of the choice (56) into (54). Then $P\leq r$
because $H_1$ is an increasing function. Hence, the minus sign must be
chosen in (54). When $u$ increases $P$ necessarily decreases, becomes null
and sometimes even negative as (57) demonstrates. However, for fixed $r$ it
remains bounded in order to keep the root in (54) real. The integral in (54)
also remains finite so there should be some limit for the growth of $u$ like
(61). The same happens in (55) if we stick to the main branch of $\arcsin x$
. Fortunately, $u\left( P\right) $ may be a multivalued function while $%
P\left( u\right) $ can not be. This explains why there are no problems in
this case. Multivalued functions resulted from the inversion of periodic
functions. In the general case periodic functions do not appear in $K$ and
that causes the limit problem. If $P$ is not extended to negative values the
limit of $u$ is also a coordinate singularity and is given by 
\begin{equation}
u=K\left( r\right) -K\left( 0\right)  \label{sixtytwo}
\end{equation}
This singularity is present generically in solutions with (56).

Another choice is 
\begin{equation}
c\left( r\right) =2H_0=2H_1\left( P_0\right)  \label{sixtythree}
\end{equation}
where $H_0$ is some very big constant. Then we may take the positive sign in
(54) and $P_0\geq P\geq r$ . The $P$ , $u$ and $r$ dependencies separate: 
\begin{equation}
K\left( P\right) =u+K\left( r\right)  \label{sixtyfour}
\end{equation}
There is no coordinate singularity but the region $P>P_0$ is not described
by this coordinate system because $K$ is ill-defined there. In turn this
means 
\begin{equation}
\begin{array}{llll}
r<P_0 &  &  & u<K\left( P_0\right) -K\left( 0\right)
\end{array}
\label{sixtyfive}
\end{equation}

For the exterior solution these inequalities look like 
\begin{equation}
\begin{array}{llll}
r<ae^{\frac{H_0}b} &  &  & u<a\sqrt{\frac \pi {2b}}e^{\frac{H_0}b}{\rm 
erf}\sqrt{\frac{H_0}b-\ln \frac ra}
\end{array}
\label{sixtysix}
\end{equation}
If the first of them is made stronger and $H_0$ is taken big enough, $u$ and 
$r$ can cover a lot of their range. The choice (63) is perfect if $H_1\left(
P\right) $ were bounded from above and $H_0>H_{1\max }$ . Unfortunately,
this does not happen due to the lower limit of the inside integral in (20).
It cures the singular behaviour at small $P$ but generates a logarithmic
term in $H_1$ like the first term in (51).

A problem arises when we try to join the interior and exterior solutions
discussed above. We start with (47) and $\delta \left( u\right) $ replaced
by $\theta \left( u\right) $ \cite{eleven}. Now we can't replace e.g. $%
\theta \left( a-P\right) $ by $\theta \left( a-r\right) $ and that makes
eq(54) intractable. Like in the case of impulsive waves it is preferable to
have one smoothly falling out $\rho $ for all $r$ like the example given by
(51). With such $H_1$ the integral in (54) can not be evaluated analytically
and the limit problem still exists. This is the best we can do for realistic
shock waves.

\section*{Acknowledgement}

This work was supported by the Bulgarian National Fund for Scientific
Research under contract F-632.

\newpage\

\end{document}